\documentclass[prb,twocolumn,showpacs,groupedaddress,preprintnumbers]{revtex4}
\usepackage{amssymb}
\usepackage{epstopdf}
\usepackage{graphicx}
\usepackage{dcolumn}
\usepackage{amsmath}
\usepackage{amsfonts}

\begin{document}
\preprint{}

\title[Short Title]{Ultrafast Critical Dynamics of Ferroelectric Phase Transition in Pb$_{1-x}$Ge$_{x}$Te}
\author{Rong Lu$^{1,2}$}
\author{Muneaki Hase$^{1,3,4}$}
\email{mhase@bk.tsukuba.ac.jp}
\author{Masahiro Kitajima$^{1,4}$}
\author{Shinichi Nakashima$^{5}$}
\author{Shunji Sugai$^{6}$}
\affiliation{$^{1}$Advanced Nano-Characterization Center, National Institute for Materials
Science, 1-2-1 Sengen, Tsukuba 305-0047, Japan}
\affiliation{$^{2}$Department of Chemistry, Renmin University of China, 59 Zhongguancun 
Avenue, Beijing, 100872, China}
\affiliation{$^{3}$PRESTO, Japan Science and Technology Agency, 4-1-8 Honcho,
Kawaguchi, Saitama 332-0012, Japan}
\affiliation{$^{4}$Institute of Applied Physics, University of Tsukuba,
Tennodai, Tsukuba, 305-8573, Japan}
\affiliation{$^{5}$Power Electronics Research Center, National Institute of Advanced
Industrial Science and Technology, 1-1-1 Umezono, Tsukuba 305-8568, Japan}
\affiliation{$^{6}$Department of Physics, Faculty of Science, Nagoya University,
Furo-cho, Chikusa-ku, Nagoya 464-8602, Japan}
\date{\today}

\begin{abstract}
Transient responses of the electronic excitation and coherent soft phonon are
investigated both above and below the ferroelectric phase transition temperature T$_{c}$ 
in Pb$_{1-x}$Ge$_{x}$Te by using an optical pump-probe technique. The coherent soft 
mode shows large red-shift and heavily over-damped decay as the temperature approaching 
T$_{c}$ from the low temperature side and the soft mode disappears above T$_{c}$.
The transient electronic response exhibits an abrupt change across T$_{c}$.
The critical behaviors of both the phononic and electronic dynamics 
are interpreted by the ferroelectric phase transition.
\end{abstract}


\maketitle 

Until now optical rewritable memory media commonly used in commercial disks 
is Ge$_{2}$Sb$_{2}$Te$_{3}$ (GST) system,\cite{Kolobov04} in which the 
laser-induced phase change between crystalline and amorphous phases 
predominates the recording process. The rapid phase transition time 
in GST was found to be several nanoseconds range. On the other hands, utilizing 
the reversal of ferroelectric domains by ultrashort optical pulses, optical data storage 
or optical switching with time scale of less than 10 picoseconds has been 
proposed.\cite{Fahy94} In this optical switch, 
one may switch the orientation of ferroelectric domains within a few phonon periods 
in time using femtosecond laser excitation, where the phonon displacement 
reaches a critical displacement of an order of 1\% of the inter-atomic 
distance.\cite{Decamp01,Hase02}  

Ferroelectric structural phase transition has been extensively investigated
by monitoring the soft mode, i.e., the lowest
frequency transverse optical (TO) phonon, whose zone center
frequency $\omega_{TO}$ is
drastically reduced toward zero near the critical point
T$_{c}$.\cite{Brinc74,Scott74} By using frequency domain techniques, such
as Raman and neutron scattering, temperature dependence of
$\omega_{TO}$ has been examined,\cite{Scott74} although the observation of 
the soft mode near T$_{c}$ is very difficult because of the extremely low 
frequency and broad linewidth of the soft mode. 

Motivated by the observation of real-time dynamics of ferroelectric 
phase transitions, a few researches have examined femtosecond 
pump-probe measurements of the coherent soft modes in {\it 
order-disorder} type perovskites, such as KNbO$_{3}$\cite{Dougherty92} 
and SrTiO$_{3}$,\cite{Kohmoto06} 
and in {\it displacive} type ferroelectric semiconductor GeTe.\cite{Hase03} 
The generation mechanism of the coherent soft mode in perovskites was 
considered to be {\it non-resonant} impulsive stimulated Raman scattering 
(NR-ISRS),\cite{Dougherty92} while in optically absorbing media, like GeTe, 
the coherent phonon can be generated by {\it resonant} impulsive stimulated Raman 
scattering (R-ISRS),\cite{Stevens02} which is recognized to 
be the general case of the displacive excitation of coherent phonons 
(DECP) mechanism.\cite{Zeiger92}
Note that those pump-probe measurements were not done at the higher temperature 
than T$_{c}$ because the soft mode becomes Raman inactive due to the 
symmetry changes and thus it will disappear at T $>$ T$_{c}$. 

Femtosecond pump-probe reflectivity technique has also been applied to 
observing coherent phonon dynamics in GST film by F\"{o}rst {\it et al}.\cite{Forst00} 
They observed drastic changes of the coherent optical phonon properties 
across the phase transition temperatures, and concluded that there was 
a metastable phase between the crystalline and amorphous phases. 
Until now, however, the time-domain study of the soft mode and electronic transient 
responses associated with the 
ferroelectric phase transition at the temperatures both above and below 
T$_{c}$ is still missing. 

In this paper, we investigate both the coherent soft phonon and 
transient electronic response in Pb$_{1-x}$Ge$_{x}$Te (PGT) utilizing 
a femtosecond optical pump-probe technique in a wide range of the lattice 
temperature. 
The ferroelectric material PGT is a narrow band-gap 
semiconductor (E$_{g}$$\sim$0.3 eV) with a high carrier mobility
and is useful for infrared laser and detectors.\cite{Suski82} It
exhibits structural phase change from rhombohedrally-distorted
structure below T$_{c}$ to cubic rocksalt structure above T$_{c}$,
and T$_{c}$ depends on the composition $x$, promising that one can
control T$_{c}$ by changing
$x$.\cite{Murase79,Hager86,Teraoka82} 
The spectroscopic measurements revealed that the band-gap energy 
in PGT is very sensitive to the ferroelectric phase
transition,\cite{Teraoka82} suggesting that ultrafast dynamics of photogenerated 
carriers exhibit significant changes across T$_{c}$.

The time-resolved reflectivity measurement was performed on a
single crystal of Pb$_{1-x}$Ge$_{x}$Te ($x$ = 0.07) obtained by
Bridgman method. At this composition $x$, T$_{c}$ is expected to
be 160$\sim$170 K.\cite{Hager86} In PGT the TO phonons ($A_{1}$
and $E$ symmetries) are soft modes, whose displacement is along
the cubic (111) direction (A$_{1}$ mode) and is perpendicular to
the cubic (111) direction ($E$ mode), and those modes are
responsible for the phase transition.\cite{Murase79} The light
source used was a mode-locked Ti:sapphire laser amplified with a
center wavelength of 800 nm (1.55 eV) and a pulse duration of
$\sim$150 fs.
The pump and probe beams were focused on the sample to a diameter
of $\sim$70 $\mu$m and the pump fluence (F$_{p}$) was reduced to
below 0.1 mJ/cm$^{2}$ to prevent heating the sample. The probe
fluence was fixed at 0.02 mJ/cm$^{2}$.
The transient isotropic reflectivity change
($\Delta R/R$) was recorded to observe the coherent A$_{1}$ mode 
as a function of the time delay at the lattice temperature (T$_{l}$) from 
8 to 200 K.

\begin{figure}[ptb]
\includegraphics[width=8.5cm]{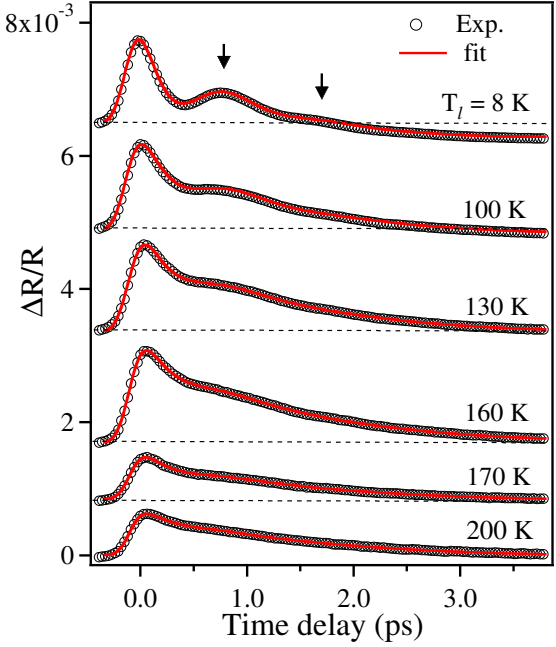}
\caption{(Color online) Transient reflectivity changes at the
different T$_{l}$ as a function of the delay time. Open circles
represent experimental data, and solid curves are the fitting using Eq. (1). }
\label{fig1}
\end{figure}
Figure 1 shows the $\Delta R/R$ signal observed at various T$_{l}$
from 8 to 200 K at the constant fluence of F$_{p}$ = 0.1 mJ/cm$^{2}$. The
fast transient signal arising from the negative time delay
represents linear electronic response with a few
picoseconds relaxation time,\cite{Note1} whereas the adjacent dip
and the following coherent oscillation (indicated by arrows in
Fig. 1.) is the contribution from the coherent lattice vibration.
The frequency of the coherent lattice vibration is $\approx$ 1.0
THz (= 33 cm$^{-1}$) at 8 K. The Raman measurements showed that
the peak frequency of the A$_1$ mode in Pb$_{1-x}$Ge$_{x}$Te with
the different composition of $x$ = 0.05 was $\approx$ 23
cm$^{-1}$.\cite{Murase79} The larger Ge composition of our
sample ($x$ = 0.07) would result in higher frequency of the A$_1$
mode,\cite{Hager86} and in addition, we are using conventional
isotropic reflectivity technique which dominates totally symmetric
A$_1$ mode rather than the E mode.\cite{Dekorsy95}  Therefore, this
coherent oscillation corresponds to the soft A$_1$ mode. It should
be emphasized that the fast electronic transient drastically changes: 
the transient electronic amplitude suddenly
decreases around 170 K as shown in Fig. 1. These features imply
that the phase transition occurs between 160 and 170 K, and this
value is quite close to the T$_{c}$ value reported by Raman
scattering at the same Ge composition.\cite{Hager86}

In order to subtract the transient electronic response from the time-domain data
and to obtain only coherent soft phonon, we utilizes the following function to fit
the data,
\begin{eqnarray}
\frac{\Delta R(t)}{R_0} = H(t)[A_{1}e^{-t/{\tau_{1}}} + A_{2}e^{-t/{\tau_{2}}} + 
Be^{-\gamma t}\cos(\omega t+\phi)]
\end{eqnarray}
where, H(t) is the Heaviside function convoluted with Gaussian to
account for the finite time resolution. A$_{1}$ and A$_{2}$ are the amplitudes, and 
$\tau_{1}$ and $\tau_{2}$ are the relaxation time of the fast and slow electronic 
responses, respectively.\cite{Note2} B, $\gamma$, $\omega$ and $\phi$ are 
respectively the amplitude, damping rate, frequency, and initial phase of the underdamped
soft A$_1$ phonon.  As shown in Fig. 1, Eq. (1) fits the data very well at all T$_{l}$.

\begin{figure}[ptb]
\includegraphics[width=8.8cm]{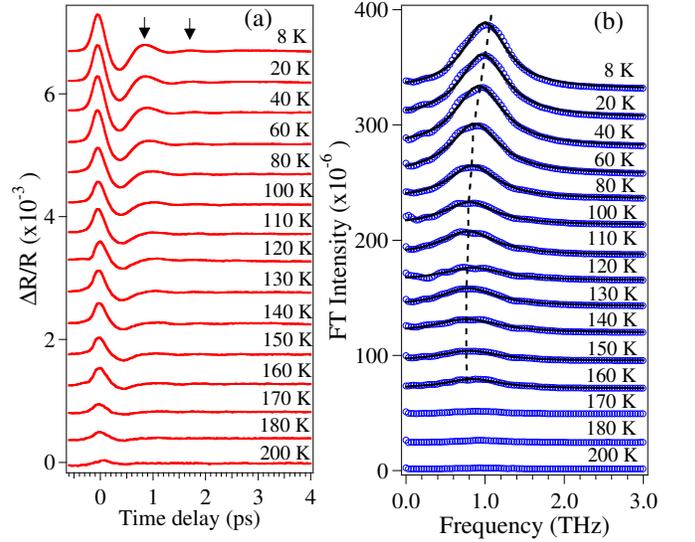}
\caption{(Color online) (a) Oscillatory part of the transient reflectivity at
different T$_{l}$. The arrows indicate coherent soft mode. (b) The FT spectra obtained
from the time-domain data in (a). Open circles
represent experimental data, and solid curves are the fitting using Lorentz function.
The dotted curve represents the peak frequency. }
\label{fig2}
\end{figure}
Figure 2(a) shows the oscillatory part of the changes of the
reflectivity, which is obtained by extracting the exponential
decay term from the above fitting at various T$_{l}$ from 8 to 200
K at the constant F$_{p}$ = 0.1 mJ/cm$^{2}$. With increasing the
T$_{l}$, the coherent A$_1$ oscillation becomes weaker and almost
vanishes at T$_{l}$ $\approx$170 K. This is explained by the fact
that the A$_1$ mode is the soft mode, whose frequency reduces
toward zero and damping becomes larger;  we can detect
only monocycle oscillation (the second cycle is rather weak )
below T$_{c}$ (= 160 $\sim$ 170 K). Figure 2(b) represents the
Fourier transformed (FT) spectra obtained from the time-domain
data in (a). In Fig. 2(b), at T$_{l}$ = 170 K, the A$_1$
mode almost disappears, supporting the idea that T$_{c}$ is 160 $\sim$ 170
K. The FT spectra show the decrease in the peak
frequency $\omega$ and the broadening of the linewidth of the coherent
A$_1$ mode (this corresponds to the damping rate $\gamma$) 
as T$_{l}$ increases as shown in Fig. 3. 
The frequency $\omega$ gradually decreases with increasing T$_{l}$, while 
the damping $\gamma$ increases until T$_{l}$ $\sim$ 140 K. 
The broadening of the linewidth of the coherent phonon spectra with T$_{l}$ 
is explained by the enhanced anharmonic phonon-phonon coupling between the
soft optical phonon and acoustic phonons,\cite{Menendez84} as 
observed in Bi.\cite{Hase98}
The T$_{l}$-dependence of $\omega$ just below T$_{c}$ significantly deviates 
from the Landau's mean field theory\cite{Landau37}; 
\begin{eqnarray}
\omega = \alpha (T_{c} -T)^{\beta}
\end{eqnarray}
which describes the second-order phase transition, where, amplitude $\alpha$ and 
exponent $\beta$ are the fitting parameters.
As a result of the best fitting using the data at T$_{l}$ $\leq$ 140 K,   
$\alpha$ =  0.47  and $\beta = 0.14 \pm 0.02$ are obtained 
for the fixed value of T$_{c}$ = 165 K. The deviation of the experimental $\omega$ 
from the Landau theory near T$_{c}$ suggests that the ferroelectric 
phase transition in PGT may be a mixture of the first-order and the second-order 
characters in the critical region due to the very large fluctuation amplitudes of the 
atoms.\cite{Muller71} 
\begin{figure}
\includegraphics[width=7.5cm]{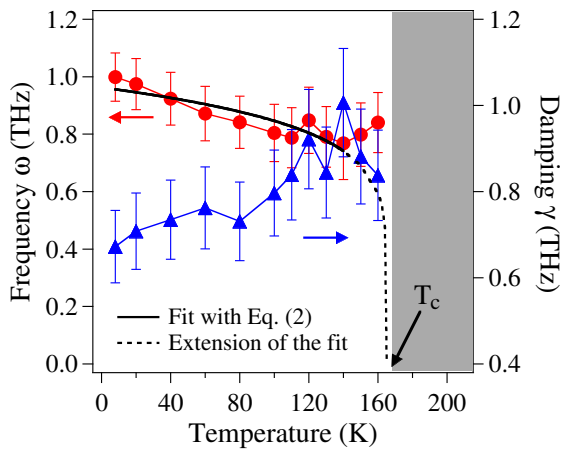}
\caption{(Color online) Temperature dependence of the frequency
$\omega$ and the damping rate $\gamma$. The closed triangles and
circles represent the experimental data. The solid curve
is the fit of $\omega$ with Eq. (2) and the dotted curve represents extension
of the fitting using Eq. (2) with the obtained parameters. The gray region 
at T$_{l}$ $\geq$ 170 K represents the region in which no soft mode available. }
\label{fig3} 
\end{figure}

In figure 4, the amplitudes of the electronic transient and that of the 
coherent A$_{1}$ mode are plotted as a function of
T$_{l}$ in order to discuss more details of the drastic change of
the electronic transient as observed in Fig. 1.\cite{Note3} At T$_{l}$ $\approx$ 
165 K the electronic amplitude ($\Delta R/R_{e}$) decreases from 
1.2$\times$10$^{-3}$ to 0.6$\times$10$^{-3}$, which corresponding to 
$\sim$ 50 \% huge
relative change. Since this huge change in $\Delta R/R_{e}$ occurs
at the same temperature of the expected T$_{c}$ value, we
interpret the drastic change in the fast electronic
transient in terms of the result of ferroelectric phase transition. On the
other hand, the A$_{1}$ mode amplitude ($I_{A1}$) gradually
decreases with T$_{l}$, and becomes almost zero at T$_{l}$ = 170
K, being consistent with the ferroelectric character that the intensity of 
the soft mode becomes weak due to strong damping below T$_{c}$ 
and it vanishes due to the symmetry change at T$_{l}$ $\geq$
T$_{c}$.\cite{Scott74}
\begin{figure}
\includegraphics[width=8.0cm]{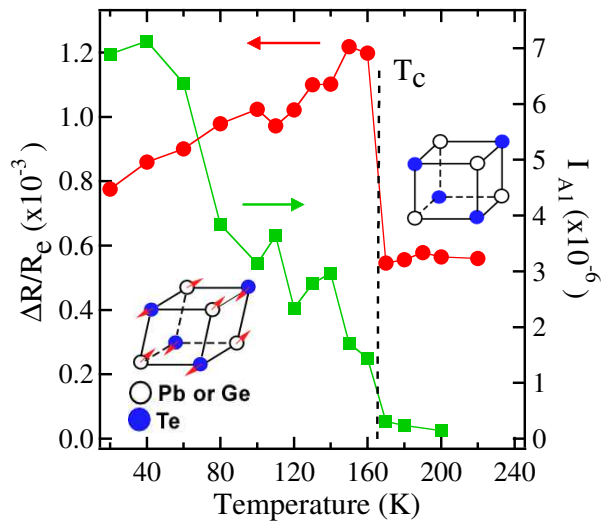}
\caption{(Color online) Temperature dependence of the
amplitude of the electronic transient (closed circles) and that of the
FT intensity of the coherent A$_{1}$ mode (closed squares) with solid
guided lines. Below T$_c$, the arrows represent the displacement of the
soft A$_{1}$ mode along (111) direction. }
\label{fig4}
\end{figure}

The structural phase transition from ferroelectric phase
(T$_{l}$$<$T$_{c}$) to paraelectric phase (T$_{l}$$>$T$_{c}$)
accompanies displacement of the ions in the unit cell and modifies
the ionic polarization: the spontaneous polarization below T$_{c}$
in the ferroelectric phase is large, while that above T$_{c}$ in
the paraelectric phase vanishes.\cite{Brinc74} This
ferroelectric phase transition then induces the changes in the
electronic band structures,\cite{Teraoka82}
including the position of the acceptor levels.\cite{Benguigui96} 
Based on the above considerations, possible origins for the drastic 
change in $\Delta R/R_{e}$ are considered as follows. 
First, the change in the electronic band structures would result in the
change in the band-gap via deformation potential (or more possibly
piezoelectric effect due to spontaneous polarization) and also the
change in the carrier mobility due to the electron-phonon
interaction.\cite{Yu99} These effects would change the interband
optical absorption coefficient, and thus may change the
electronic response $\Delta R/R_{e}$. Second, the change in
the position of the acceptor levels affects the free carrier
concentration,\cite{Benguigui96} and this effect will result in the
modification of the free carrier absorption, which contributes to the
electronic response $\Delta R/R_{e}$. By using the huge change in the 
electronic transient $\Delta R/R_{e}$ across T$_{c}$, optical switching 
or optical memory using PGT can be proposed. In this case,  
photo-induced phase transition between ferroelectric and paraelectric 
phases is required. 
Such the photo-induced phase transition will be realized if one could 
precisely control both the amplitude\cite{Hase96,Iwai06} and the 
frequency\cite{Hase03} of the coherent soft phonon. 

In summary, we have investigated ultrafast dynamics of both the coherent
soft mode and the transient electronic polarization in ferroelectric semiconductor
below and above T$_{c}$ by using the pump-probe technique. Low frequency
coherent soft mode in PGT was observed only below T$_{c}$,
showing the red-shift of the frequency and heavily over-damped behavior as the
temperature approaches T$_{c}$. The drastic decrease in the transient electronic response was
revealed across T$_{c}$, suggesting the change in the electronic
band structure, including the position of the acceptor levels.

The authors acknowledge J. Demsar and O. V. Misochko for helpful comments. 
This work was supported in part by a Grant-in-Aid for the Scientific Research from MEXT of 
Japan under grant KAKENHI -16032218, and by Iketani Science Technology Foundation 
of Japan.

\end{document}